\documentstyle[epsf,10pt,aas2pp4]{article}

\def\kms{\relax \ifmmode {\,\rm km\,s}^{-1}\else \,km\,s$^{-1}$\fi}
\def\ha{\relax \ifmmode {\rm H}\alpha\else H$\alpha$\fi}
\def\hb{\relax \ifmmode {\rm H}\beta\else H$\beta$\fi}
\def\hi{\relax \ifmmode {\rm H\,{\sc i}}\else H\,{\sc i}\fi}
\def\hii{\relax \ifmmode {\rm H\,{\sc ii}}\else H\,{\sc ii}\fi}
\def\h2{\relax \ifmmode {\rm H}_2\else H$_2$\fi}
\def\lha{\relax \ifmmode L_{{\rm H}\alpha}\else $L_{{\rm H}\alpha}$\fi}
\def\shi{\relax \ifmmode \sigma_{{\rm HI}}\else $\sigma_{\rm HI}$\fi}
\def\sh2{\relax \ifmmode \sigma_{{\rm H}_2}\else $\sigma_{{\rm H}_2}$\fi}
\def\degr{\hbox{$^\circ$}}
\def\arcmin{\hbox{$^\prime$}}
\def\arcsec{\hbox{$^{\prime\prime}$}}
\def\deg{\hbox{$^\circ$}}

\def\fdg{\hbox{$.\!\!^\circ$}}
\def\fs{\hbox{$.\!\!^{\rm s}$}}
\def\farcm{\hbox{$.\mkern-4mu^\prime$}}
\def\farcs{\hbox{$.\!\!^{\prime\prime}$}}
\def\degd#1.#2{ #1\fdg#2 }                 
\def\mind#1.#2{ #1\farcm#2 }               
\def\secd#1.#2{ #1\farcs#2 }               
\def\hhh{\ifmmode {\rm ^h}              
         \else {${\rm ^h}$}
         \fi}
\def\sss{\ifmmode {\rm ^s}              
         \else {${\rm ^s}$}
         \fi}
\def\hms#1h#2m#3s{                      
                  \relax
                  \ifmmode #1^{\rm h}\,#2^{\rm m}\,#3^{\rm s}
                  \else \hbox{$#1^{\rm h}\,#2^{\rm m}\,#3^{\rm s}$}
                  \fi
                 }
\def\dms#1d#2m#3s{                      
                  \relax
                  #1\degr\,#2\arcmin\,#3\arcsec 
                 }
\def\hmsd#1h#2m#3.#4s{                  
                      \relax
                      \ifmmode #1^{\rm h}\,#2^{\rm m}\,#3\fs#4
                      \else \hbox{$#1^{\rm h}\,#2^{\rm m}\,#3\fs#4$}
                      \fi
                     }
\def\dmsd#1d#2m#3.#4s{                  
                      \relax
                      #1\degr\,#2\arcmin\,#3\farcs#4
                     }
\def\mag{\relax                          
        \ifmmode ^{\rm m}
        \else $^{\rm m}$
        \fi
       }
\def\magd#1.#2{                          
              \relax
              \ifmmode #1^{\rm m}
                       \hskip-0.55em.\hskip0.22em#2
              \else \hbox{#1$^{\rm m}
                    \hskip-0.55em.\hskip0.22em$#2}
              \fi
             }

\begin{document}
\title{Resolved structure in the nuclear region  of the ultraluminous 
infrared galaxy Mrk~273}

\author{J.H. Knapen, S. Laine, J.A. Yates, A. Robinson}
\affil{Department of Physical Sciences, University of Hertfordshire,\\
Hatfield, Herts AL10 9AB, UK}
\authoremail{knapen, seppo, jyates, ar@star.herts.ac.uk}

\author{A.M.S. Richards}
\affil{Nuffield Radio Astronomy Laboratories, University of Manchester,\\
Jodrell Bank, Macclesfield, Cheshire SK11 9DL, UK}
\authoremail{amsr@jb.man.ac.uk}

\author{R. Doyon, D. Nadeau}
\affil{Observatoire du Mont M\'egantic and
D\'epartement de Physique,
Universit\'e de Montr\'eal,
C.P. 6128, Succursale Centre Ville,   
Montr\'eal (Qu\'ebec), H3C 3J7 Canada}
\authoremail{doyon, nadeau@astro.umontreal.ca}

\begin{abstract}

We have studied the core morphology of the ultraluminous infrared
galaxy Mrk~273 by combining a high-resolution adaptive optics
near-infrared image with an optical image from the Hubble Space
Telescope and interferometric radio continuum data, all at spatial
resolutions of 150 mas or better.  The near-infrared image reveals
that the nucleus has two main components, both of which have radio
counterparts. The strongest component (N) shows very similar extended
structure in the radio and near-infrared. It has a flat radio spectrum
and is resolved into a double-lobed structure (Ne; Nw), with a
separation of $90\pm5$ mas (70~parsec).  A similar structure is
detected in the near-infrared.  We identify this component as the
location of the active nucleus.  The second component (SW), strong in
the near-infrared but relatively weak in the radio, is located $\sim1$
arcsecond to the southwest.  We interpret this as an obscured
starburst region associated with the merger.  The radio continuum
images show a third, strong, component (SE) which has previously been
interpreted as a second nucleus.  However, it shows no associated
optical or near-infrared emission, suggesting that it is in fact a
background source.

\end{abstract}

\keywords{Galaxies: Active -- Galaxies: Individual: Markarian 273 
-- Galaxies: Interactions -- Galaxies: Nuclei -- Galaxies: Seyfert --
Radio Continuum: Galaxies}

\newpage

\section{Introduction}

Ultraluminous infrared galaxies (ULIRGs) are among the brightest known
galaxies, some with  $L_{bol}\ge 10^{12}\,L_{\odot}$, comparable
with quasars.  These extreme, mostly infrared,  luminosities are
thought to be produced by dust which enshrouds, and is heated by, a
powerful source of optical--UV continuum.   The most likely
candidates for this energy source are an active galactic nucleus (AGN)
or a massive starburst (Joseph \& Wright 1985; Sanders et al.  1988;
Sanders, Scoville \& Soifer 1991).  In either case, it is probable
that the nuclear activity is triggered by the merger of two gas-rich
galaxies (Solomon et al. 1997).  Optical imaging shows that ULIRGs
generally exhibit morphological peculiarities such as tidal tails or
double nuclei which are often associated with galaxy mergers (e.g.
Sanders et al.  1988; Melnick \& Mirabel 1990).

Mrk~273 (UGC 08696; 1342+561), at a distance of 151~Mpc\footnote{We
adopt $H_0=75\,$km\,s$^{-1}$\,Mpc$^{-1}$.  At the distance of Mrk~273
one arcsecond corresponds to 0.7\,kpc.}, is one of the most luminous
and best studied IRAS galaxies, with $\log\,(L_{FIR}/L_\odot)=12.04$.
Optical spectroscopy reveals strong emission lines 
characteristic of a Seyfert 2 nucleus (Koski 1978; Sanders et al.
1988), although the high Balmer decrement suggests that the narrow
emission line region suffers considerable extinction ($A_V \sim3$
mag).  On the other hand, Goldader et al.  (1995) find evidence for a
starburst signature in a 3 arcsec aperture $K$-band spectrum.  The
optical morphology of Mrk~273 suggests a highly disturbed system, the
most striking feature being a ``tail'' extending nearly an arcminute
($\sim40$~kpc) south of the main body of the galaxy (Fig.~1), strongly
indicative of a merger remnant (cf. Toomre \& Toomre 1972).  More
detailed imaging observations suggest that the latter may have a
multiple nucleus (Koroviakovskii et al.  1981; Mazzarella \& Boroson
1993).  Near-infrared (NIR) imaging by Majewski et al. (1993) first
showed a double structure at 2.2 $\mu$m, with components to the north
and southwest separated by $\sim1.1$ arcsec.  Evidence for a multiple
nucleus has also been presented by Condon et al.  (1991), who mapped
Mrk~273 as part of an 8.44\,GHz survey of ULIRGs.  Their 300 mas
resolution image shows two strong radio components (N and SE)
separated by 800 mas which they interpret as a double nucleus
resulting from a recent merger.

New developments in optical, NIR and radio
interferometry imaging technology have dramatically increased the
spatial resolution at which multi-wavelength morphological studies can
be performed.  We have used these techniques to study the core
structure of Mrk\,273 with the aim of testing the merger hypothesis
for this object, and, by extension, other ULIRGs.  In this letter, we
present a data set consisting of NIR $K$-band, optical 7940 \AA\, and
1.6, 5 and 8.4 GHz radio continuum images, all at resolutions around,
or better than, 150 mas. 

\section{Observations and data processing}

\begin{figure*}

\epsfxsize=17cm \epsfbox{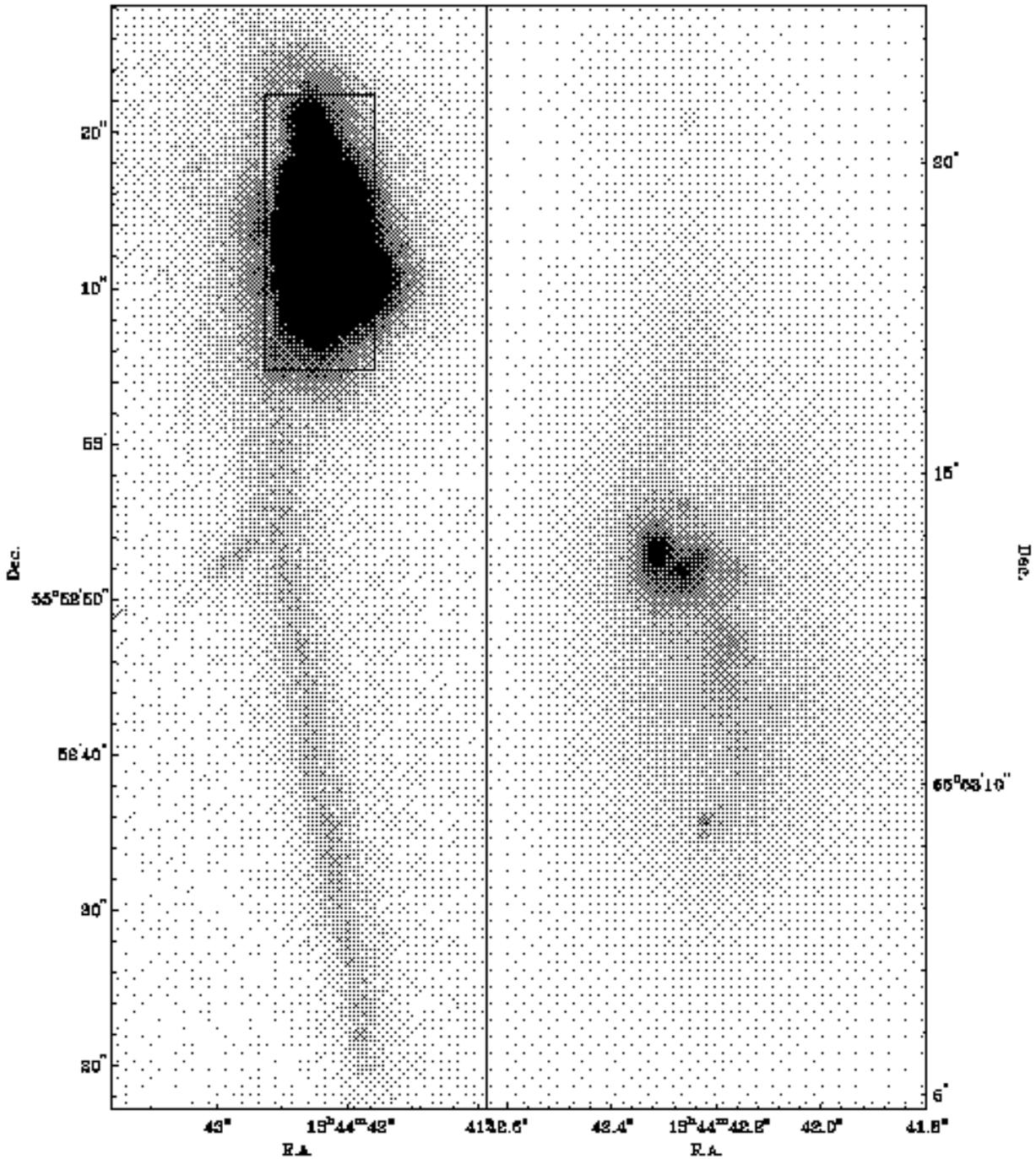}
\small{
\caption{Archival HST image of Mrk 273, centered at
$\lambda=7940$\,\AA. The galaxy is shown on the left, revealing the
long tidal tail towards the south. The inset on the right shows the
nuclear region. 1 arcsec corresponds to $\sim70$~pc. \label{fig1}}
}
\end{figure*}

The optical image of Mrk~273 (Fig.~1) was obtained from the HST
archive. It is a WFPC2 image taken through the F814W ($I$-band)
filter. We obtained the flat-fielded image and removed cosmic ray hits
interactively (a few remain in the nuclear region). 

\begin{figure*}
\epsfxsize=14cm \epsfbox{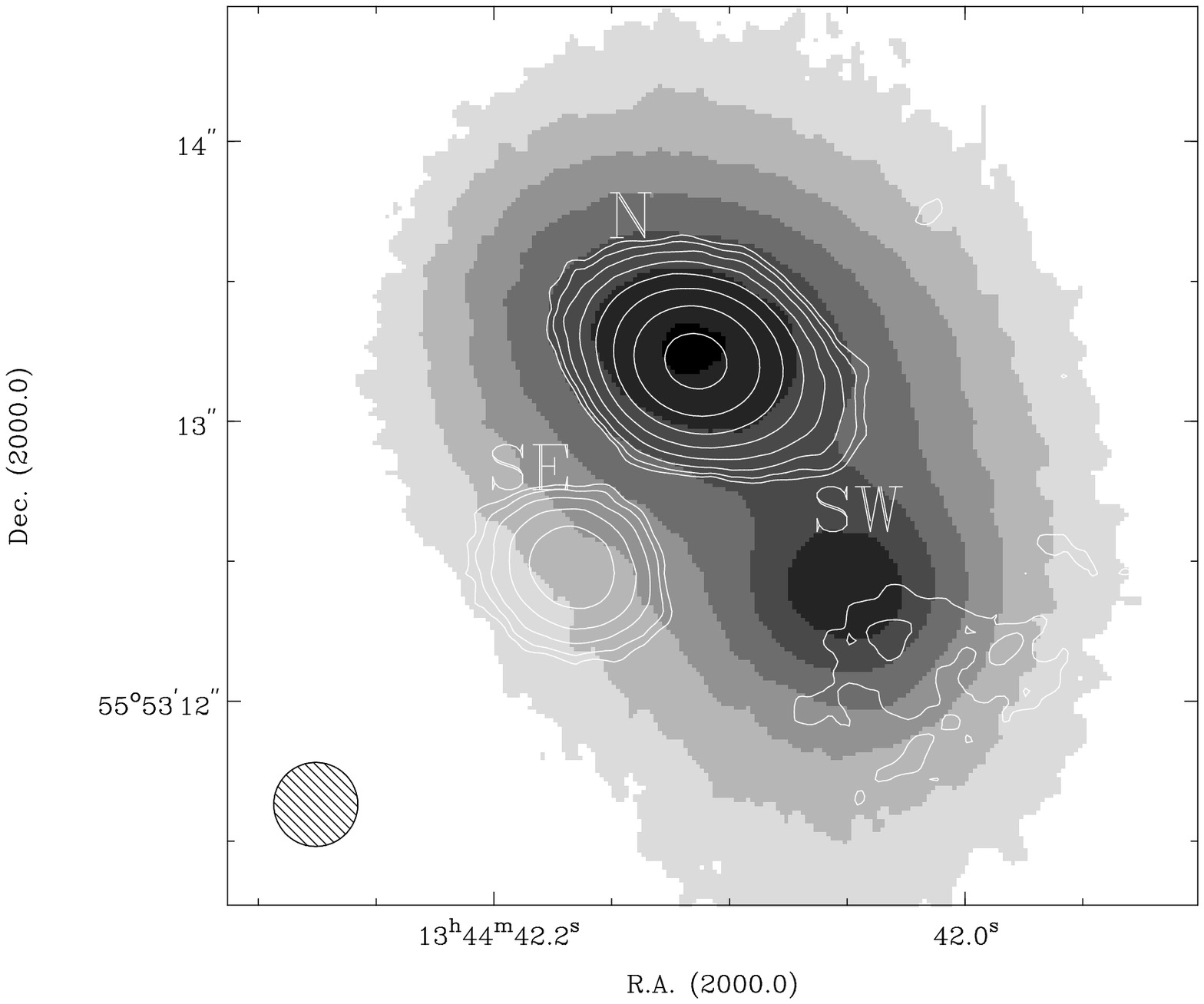}
\small{
\caption{Contours of the 300 mas
resolution MERLIN 5~GHz radio image of the nucleus of Mrk 273 overlaid
on the 150 mas resolution CFHT $K$ band image (greyscale).  The three
different components have been labelled N, SE and SW.  The beam size
of the radio image is shown at the bottom left.  The contour levels of
the 5~GHz radio image are $9.0\times10^{-5}\times$ (2 [=2$\sigma$], 4,
8, 16, 32, 64, 128 and 256) Jy/beam, and the greyscale levels of the
$K$ band image 15.0 (light), 14.6, 14.2, 13.8, 13.4, 13.0 and 12.6
(dark) mag arcsec$^{-2}$ \label{fig2}}
}
\end{figure*}

The $K$-band image (Fig.~2) was obtained on the night of 19 March 1997
on the Canada-France-Hawaii Telescope (CFHT) using the adaptive optics
system PUEO (Rigaut et al. 1997a), combined with the NIR camera MONICA
(Nadeau et al. 1994). PUEO  uses a curvature mirror and a wavefront
sensor working  with guide stars up to $R$ magnitude 14 to give
diffraction limited images at $H$ and $K$.  Individual images were
sky-subtracted, flatfielded, corrected for bad pixels and
combined in a final mosaic. The uncorrected seeing was $\sim$600 mas,
while the corrected resolution was near the diffraction limit at $K$,
125 mas. We used throughout the image scale of 34.38$\pm$0.07
mas/pixel and rotation prescription from Rigaut et al. (1997b). The
orientation of the $K$-band image is accurate to within 0.1 degree.

\begin{table*}
\caption{Summary of radio continuum data.  $S_P$ is peak flux, $S_I$ is
integrated flux $>5\sigma$ over the area given.  The integrated flux
error is less than 0.2 mJy.\label{table1}}
\begin{center}
{\small
\begin{tabular}{cccccccccccc}
$f$   & Beam&\multicolumn{3}{c}{$S_P$}&
\multicolumn{3}{c}{$S_I$}&\multicolumn{3}{c}{Area $>5\sigma$} 	& rms \\
(GHz) &(mas)	&\multicolumn{3}{c}{(mJy/beam)}	&
\multicolumn{3}{c}{(mJy)}	    &\multicolumn{3}{c}{(arcsec$^{2}$)}	& 
(mJy/bm)\\
\tableline
  	&    	& N   &  SE   &   SW  		& N   &  SE   &   
SW  		    & N   &  SE   &   SW      			&   \\
\tableline
1.6 & 150       &17.9&11.4 &$<$0.5 		
&71.7&17.4&		    	&0.340&0.131&			& 0.15 \\
1.6 & 300       &41.0&15.9 &0.6   			
&&&	                &$\pm0.031$&$\pm0.019$&			& 0.15 \\
&&&&&&&&&&&\\
5 &   50        &4.8\tablenotemark{a} &2.4  &$<$0.1 		&40.8&4.8&			
&0.145&0.023&		& 0.04 \\
5 &   150       &17.0&2.4  &$<$0.3 			&&&			&$\pm0.006$&$\pm0.008$&		
& 0.10 \\
5 &   300       &29.0&5.2  &0.5   		&&&		&&&								
& 0.09 \\
&&&&&&&&&&&\\
8.4 & 300       &20.4& 3.2 & 0.3   		&36.4&3.7&0.2			
&1.203&0.398&0.123				& 0.03 \\
&&&&&&&&$\pm0.183$&$\pm0.111$&$\pm0.062$&\\

\end{tabular}
\tablenotetext{a}{Peak flux of the strongest of twin peaks (Ne).}
}	

\end{center}
\end{table*}

\begin{table}
\caption{Accurate positions and $K$ photometry of the various components.
Radio continuum peak positions are from the 50 mas resolution 5 GHz
data for components Ne (error $\sim1$ mas), Nw ($\sim1$) and SE
($\sim2$), and from the 8.4 GHz 300 mas resolution data for N
($\sim2$) and SW ($\sim27$).\label{table2}}

{\small
\begin{tabular}{lcccc}
 &R.A.&Dec.&$K$&$K$\\
 &&&(mag)&(L$_{\sun}$)\\
\tableline
Ne&13$^{h}$44$^{m}$42$\fs 1192$&55$\arcdeg$53$\arcmin$13$\farcs 219$&
17.1$\pm0.2$&$7.62 \times 10^{8}$\\
Nw&13$^{h}$44$^{m}$42$\fs 1093$&55$\arcdeg$53$\arcmin$13$\farcs 185$&
19.1$\pm0.2$&$1.21 \times 10^{8}$\\
N&13$^{h}$44$^{m}$42$\fs 1171$&55$\arcdeg$53$\arcmin$13$\farcs 
182$&13.4$\pm0.2$&$2.30 \times 10^{10}$\\
SW&13$^{h}$44$^{m}$42$\fs 0372$&55$\arcdeg$53$\arcmin$12$\farcs 
144$&14.2$\pm0.2$&$1.10 \times 10^{10}$\\
SE&13$^{h}$44$^{m}$42$\fs 1677$&55$\arcdeg$53$\arcmin$12$\farcs 496$&-&-\\
\end{tabular}
}
\end{table}

Mrk 273 was observed at 4993 MHz (6 cm, 27 November 1992) and 1656 MHz
(18 cm, 13 December 1995) with the Multi-Element Radio-Linked
Interferometer Network (MERLIN).  The nearby QSO 1335+552 was used to
correct phase and amplitude errors, while OQ208 and 3C286 were used as
flux calibrators.  The maximum baseline is 217 km, giving a natural
beam size of 55 mas at 5 GHz and 170 mas at 1.6 GHz.  The bandwidths
used were 16 MHz at 5 GHz and 8 MHz at 1.6 GHz.  Noise characteristics
and peak fluxes of the radio data are shown in Table~1, and radio
positions in Table~2.  The MERLIN
positions are absolute and accurate to 2 mas at 5 GHz and 5 mas at 1.6
GHz.  These radio data can therefore be used as an astrometric
reference. The flux calibration is accurate to $2-4$\%.

The 8.44~GHz (3.6 cm) VLA radio image has been published by Condon et
al. (1991), and was obtained from the NED database. The size of the
restoring beam is 300~mas.  The axis defined by components N and SW
(see below)in the $K$ image coincides with that in the 8.4 GHz map. We
translated the NIR image so that its main peak coincides with the
radio component N. We show the 5~GHz radio map overlaid on the $K$
image in Fig.~2.

\section{Central morphology of Mrk~273}

On scales of 5--10 arcseconds ($\sim5$ kpc), the optical morphology is
complicated by prominent dust lanes, and several distinct regions of
strong emission, the two most prominent of which (along a
north-southwest line) are separated by $\sim2$ arcsec (Fig.~1).  The
relatively low astrometric precision of HST ($\sim0.5$ arcsec)
precludes accurate registration  with our other data sets. Nor
is it possible to accurately match strong features in the optical and
NIR, probably because the optical data are much more severely confused
by patchy dust extinction.

\begin{figure*}
\epsfxsize=12cm \epsfbox{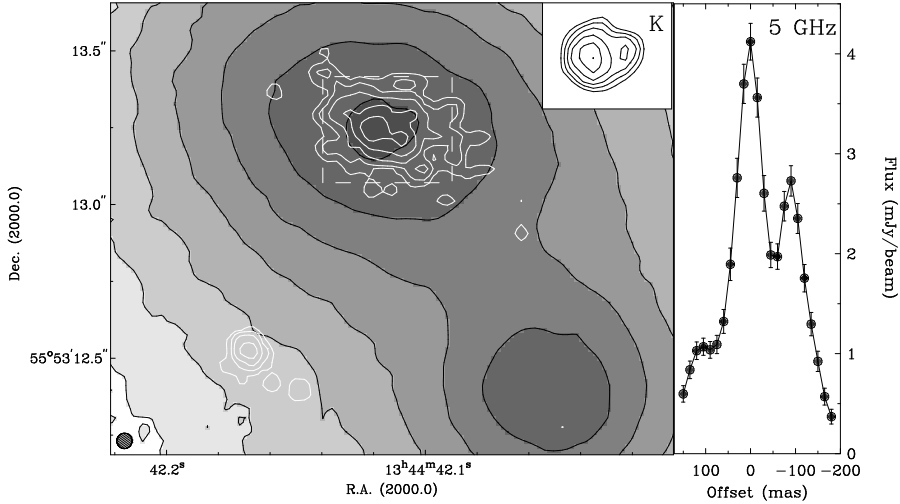}
\small{
\caption{Contours of the 50 mas resolution MERLIN 5 GHz 
radio image (white) overlaid on the 150 mas resolution CFHT $K$ band
image of the nuclear region of Mrk 273 (left).  Contour levels of the
radio image are $4.0\times10^{-5}\times$ (4 [=4$\sigma$], 8, 16, 32
and 64) Jy/beam.  The beam size is shown at the bottom left. The
greyscale levels and contours of the $K$ image are as in Fig.~2.  The
inset shows, at the same scale, the twin peaks region of component N
as seen in $K$, contour levels are 12.63, 12.61, 12.59, 12.57, 12.55
and 12.53 mag arcsec$^{-2}$. Right panel shows a slice taken through
the twin peaks in N in the 50 mas 5 GHz MERLIN map: a line 330 mas
long at position angle 250\deg\ from $\alpha=$\hmsd 13h44m42.1359s,
$\delta=$\dmsd 55d53m13.270s. The error-weighted mean of the flux was
taken in strips 75 mas deep at 15 mas intervals along the line. This
flux and $3\sigma$ errors are shown.
\label{fig3}}
}
\end{figure*}

The morphology of the central kpc region of Mrk~273 is shown in
Fig.~2.  The three dominant sources of emission have been labeled N
(north), SE (southeast), and SW.  Component N is the brightest radio
and NIR peak.  It is surrounded by an elongated structure whose
extent, position angle, and ellipticity are very similar in NIR and
radio (Fig.~2).   At 50 mas resolution (5 GHz, Fig.~3), the outer
fringes of component N are  resolved into several localized sources
of emission.

At 50 mas resolution at 5 GHz, the center of component N is resolved
into two peaks, which we call Ne (east) and Nw (west)  (Fig.~3).
These ``twin peaks'' are separated by $90\pm5$ mas ($\sim$70 pc) 
spatially (Fig.~3, right).  The signal to noise ratio of the twin
peaks gives a positional accuracy of a few mas,  so they are a
robust detection.  The $K$ image also shows a double-peaked feature
(inset in Fig.~3), at the same position but with a slightly different
($\sim15\deg$) position angle. This double $K$-peak has been confirmed
in a Keck image (E.Ye. Khachikian, private communication).  

Component SE is a strong radio continuum source  (Condon et
al. 1991). We detect it clearly in our two independent MERLIN data
sets, but it has no counterpart in NIR or optical images (Figs.~1, 2,
3).  We do not confirm the tentative detection at 2.2 $\mu$m by
Majewski et al. (1993).  Fig.~3 shows that at 50 mas resolution
component SE is resolved into a strong, peaked, component with two
separate regions of emission at 4$\sigma$ level.  Component SW is
the second strong NIR source, almost 1 arcsec (700 pc) southwest of
the main emission region, N. It may have a counterpart in the optical
image, possibly distorted or displaced due to the effects of dust
extinction, but the astrometry of the optical image is not accurate
enough to confirm this. Component SW is extended and relatively weak in
both the 5 GHz (Fig.~2) and 8.4 GHz maps (Fig.~1 of Condon et
al. 1991).

\section{Fluxes and spectral indices}

\begin{figure}
\epsfxsize=8cm \epsfbox{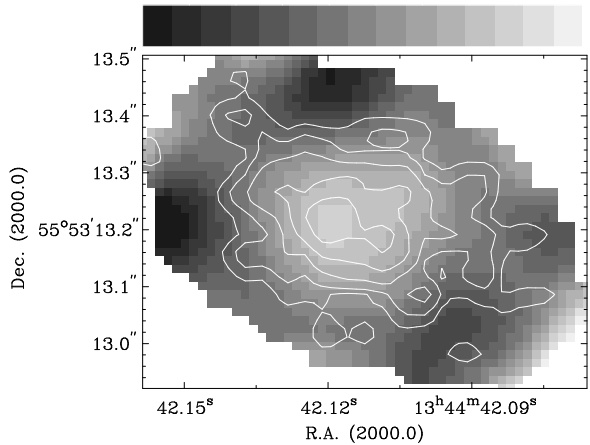}
\small{
\caption{Contours of the 50 mas resolution 5 GHz radio
image (white) overlaid on the 150 mas resolution spectral index map of
the 1.6 GHz and 5 GHz radio images.  Contour levels for the 5 GHz map
are as in Fig.~3.  Grey levels indicate spectral indices from $-1.3$
(black) to 0.3 (white) in steps of 0.1.\label{fig4}}
}
\end{figure}

Peak and integrated radio fluxes for the different components are
shown in Table~1, and radio positions and $K$-band magnitudes and
luminosities in Table~2.  We have determined spectral indices from our
combined radio continuum images, and confirmed these with direct
derivations from the peak fluxes.  Comparing the 1.6 and 5 GHz images,
both at 150 mas resolution, gives a spectral index ($\alpha$;
$S\propto\nu^\alpha$) for the twin peaks area in component N of
$-0.1\pm0.3$ (Ne and Nw are not resolved at 150 mas).  The surrounding
emission has a significantly different spectral index of $-$0.7 $\pm$
0.1 (Fig.~4).  The radio spectral index of component SE is $-$1.0
$\pm$ 0.2.  The diffuse radio emission detected at 1.6, 5 and 8.4 GHz
for component SW has a spectral index of $-0.4\pm0.5$.  The flat
spectrum in the twin peaks region in component N suggests that it may
be the compact synchrotron core of a Seyfert radio source.  This is
consistent with N being the strongest radio and NIR component.

A flat spectrum would also be expected if the radio emission is
dominated by nebular free-free emission, the strongest source of which
is likely to be the Seyfert narrow line region.  Taking the H$\beta$
flux measured by Koski (1978) and correcting it for the extinction
($A_{V}\approx 3.1$ mag) inferred from the Balmer decrement, we
estimate that for a typical nebular temperature of $T=10^4$\,K, the
free-free continuum emitted by the narrow line region can contribute
at most 1\% of the measured 5\,GHz flux density of component N.  If a
dusty molecular torus is present (as we suggest in Section 5), another
possibility is that some fraction of the observed radio flux is due to
free-free continuum emission from a hidden emission line region.
However, it is unlikely that all of the radio emission from component
N can be explained in this way, since the corresponding H$\beta$
luminosity ($\sim 3\times 10^{43}$\,erg\,s$^{-1}$) would suggest a
quasar broad-line region and such a source would be unresolved at the
spatial resolution of our radio maps.


The spectral index of the emission surrounding the twin peaks is
significantly steeper (Fig.~4). Given the morphology of this region,
the radio emission may trace supernova remnants in regions of star
formation around the AGN, at a radius of a few hundred pc.

The origin of both components SE and SW is unclear.  SE was
interpreted by Condon et al.  (1991) as the nucleus of a merging
galaxy.  However, given the absence of an NIR or optical counterpart
and its negative spectral index, which suggests optically thin
synchrotron emission,  component SE may well be a background source,
presumably a distant radio galaxy.

The flat spectrum and diffuse morphology of the relatively weak radio
source corresponding to component SW suggests optically thin free-free
emission.  The integrated 8.44\,GHz flux density would then imply an
H$\beta$ flux comparable with that measured by Koski (1978).  It is
plausible, therefore, that component SW is the site of a starburst
triggered by the merger, with the observed radio flux arising as
free-free emission from the associated \hii\ regions.  One might then
expect the emission line spectrum to exhibit a mixture of Seyfert and
starburst characteristics, but the HST image (Fig.  1) suggests that
component SW is at least partially obscured and hence line emission
from the starburst may suffer heavy extinction.  Indeed, while there
is little evidence for spectral features related to star formation in
the optical spectrum, such features do become apparent in the NIR
(Goldader et al.  1995).

\section{Discussion and summary}

We have combined state-of-the-art radio, NIR and optical imaging
observations, all at resolutions of around 150 mas, to study the
nuclear region of the ULIRG Mrk\,273.   Our images have revealed
three main radio components, two of which, N and SW, have bright NIR
counterparts.  Small-scale ($\leq150$\,mas) structure is also present
in component N and, possibly, SE.  Mrk 273 has
a double nucleus which is detected at both NIR and radio wavelengths
and whose components (N and SW) are separated by $\sim 1$ arcsec,
supporting the hypothesis that this system formed in a recent
merger. Condon et al.  (1991) reached a similar conclusion on the
basis of their radio data, but identified radio components N and SE as
the double nucleus. The lack of a NIR counterpart to component SE,
however, suggests that it is in fact a background source unrelated to
Mrk~273.

Component N is probably the site of the active galactic nucleus which
produces the Seyfert 2 emission line spectrum: it is the brightest
source at both radio and IR wavelengths, and the radio continuum,
which has a flat spectral index consistent with a compact synchrotron
source, is too strong to be attributed to nebular free-free emission.
At a resolution of 50 mas, the radio source within component N
resolves into a twin-peaked structure whose ``lobes'' are separated by
$\sim 90$ mas ($\sim 70$\,pc).  A similar double structure is detected
in our $K$-band image.  The two peaks Ne and Nw could represent
distinct merger remnants each harbouring a compact synchroton radio
source. $K$ magnitudes for components N and SW, and even Ne and Nw
(Table~2), are consistent with those of bright galactic nuclei (e.g.,
Forbes et al. 1992). However, we suggest the alternative
interpretation that the double NIR structure is actually an artifact
of obscuration by an edge-on dusty molecular torus.  The corresponding
radio structure can then be explained as a compact double-lobed radio
source similar to those present in other Seyfert galaxies, which is
fed by synchrotron-emitting plasma ejected from the (hidden) active
nucleus.  The presence and orientation of the torus should be
confirmed by forthcoming MERLIN observations of the OH megamaser
emission associated with Mrk~273.

The radio flux from component SW is relatively weak and can plausibly
be attributed to nebular free-free emission. This leads us to propose
that component SW harbors a heavily obscured starburst which may have
been triggered by the merger.

\acknowledgments

We thank Drs P.  Thomasson, T.W.B.  Muxlow and S.T.  Garrington for
assistance with the MERLIN observations, Dr.  E.Ye.  Khachikian for
helpful discussions, and Drs.  K.  Weiler and P.D.  Barthel for
comments on an earlier version of the manuscript.  MERLIN is operated
by the University of Manchester on behalf of PPARC.  JHK and AR thank
the British Council, NWO and the Royal Society for financial support.
The CFHT is operated by the NRC of Canada, the CNRS de France, and the
University of Hawaii.  The NASA/IPAC Extragalactic Database (NED) is
operated by JPL, CalTech, under contract with NASA.  Partly based on
observations made with the NASA/ESA HST, obtained from the data
archive at STScI.  STScI is operated by AURA, Inc.  under NASA
contract NAS 5-26555.


\begin{references} 

\reference{cond91}Condon, J.J., Huang, Z.-P., Yin, Q.F., Thuan, T.X. 1991, 
ApJ 378, 65


\reference{forbes}Forbes, D. A., Ward, M. J., DePoy, D. L., Boisson, C., 
Smith, M. S. 1992, MNRAS 254, 509

\reference{gold}Goldader, J.D., Joseph, R.D., Doyon, R., Sanders, D. B. 
1995, ApJ 444, 97

\reference{joseph}Joseph, R.D., Wright, G.S., 1985, MNRAS 214, 87

\reference{koro}Koroviakovskii, Y.P., Petrosian, A.R., Saakian, K.A,
Khachikian, E.Ye. 1981, Astrofizika 17, 231

\reference{koski}Koski, A.T. 1978, ApJ 223, 56

\reference{majewski} Majewski, S.R., Hereld, M., Koo, D.C., Illingworth, 
G.D., Heckman, T.M. 1993, ApJ 402, 125

\reference{mazz}Mazzarella, J.M., Boroson, T.A. 1993, ApJS 85, 27

\reference{meln90}Melnick, J., Mirabel, I.F. 1990, A\&A 231, L19

\reference{nadeau}Nadeau, D., Murphy, D.C., Doyon, R. \& Rowlands, N. 1994,
PASP, 106, 909

\reference{rigaut1}Rigaut, F., et al. 1997a, PASP, submitted

\reference{rigaut2}Rigaut, F., Doyon, R., Davidge, T., Crampton, D., 
Rouan, D., \& Nadeau, D. 1997b, Submitted to ApJ Letters

\reference{sand88}Sanders, D.B., Soifer, B.T., Elias, J.H., Madore, B.F., 
Matthews, K., Neugebauer, G., Scoville, N.Z. 1988, ApJ 325, 74

\reference{sand91}Sanders, D.B., Scoville, N.Z., Soifer, B.T. 1991, 
ApJ 370, 158

\reference{solomon97}Solomon, P.M., Downes, D., Radford, S.J.E., 
Barrett, J.W. 1997, ApJ 478, 144

\reference{toomre}Toomre, A. \& Toomre, J. 1972, ApJ 178, 623

\end{references}
\end{document}